\documentclass[a4paper,fleqn,usenatbib]{mnras}

\usepackage{newtxtext,newtxmath}
\usepackage[T1]{fontenc}
\usepackage{ae,aecompl}

\usepackage{graphicx}	
\usepackage{amsmath}	
\usepackage{amssymb}	

\def\pasp{PASP} 

\title[The structure behind the Galactic bar ]{The structure behind the Galactic bar traced by red clump stars in the VVV survey. \thanks{Based on observations collected at the ESO La Silla-Paranal Observatory 179.B-2002.}}

\author[Oscar A. Gonzalez]{
Oscar A. Gonzalez$^{1}$\thanks{E-mail:oscar.gonzalez@stfc.ac.uk},
Dante Minniti$^{2,9,10}$,
Elena Valenti$^3$,
Javier Alonso-Garc\'{\i}a$^{4,9}$,\newauthor
~Victor P. Debattista$^5$,
Manuela Zoccali$^{6,9}$, 
Marina Rejkuba$^3$,
Bruno Dias$^{7,2}$,\newauthor
~Francisco Surot $^3$,
Maren Hempel$^6$,
Roberto K. Saito$^8$
\\
  $^1$ UK Astronomy Technology Centre, Royal Observatory, Blackford Hill, Edinburgh, EH9 3HJ, UK\\
  $^2$ Departamento de F\'isica, Facultad de Ciencias Exactas, Universidad Andr\'es Bello, Av. Fernandez Concha 700, Las Condes, Santiago, Chile \\
  $^3$ European Southern Observatory, Karl-Schwarzschild Strasse 2, D-85748 Garching, Germany \\
  $^4$ Unidad de Astronom\'ia, Facultad Cs. B\'asicas, Universidad de Antofagasta, Avda. U. de Antofagasta 02800, Antofagasta, Chile \\
  $^5$ Jeremiah Horrocks Institute, University of Central Lancashire, Preston, PR1 2HE, UK \\
  $^6$ Instituto de Astrof\'{\i}sica, Facultad de F\'{\i}sica, Pontificia Universidad Cat\'olica de Chile, Av. Vicu\~na Mackenna 4860, Santiago 22, Chile\\
  $^7$ European Southern Observatory, Alonso de Cordova 3107, Vitacura, Santiago, Chile\\
  $^8$ Departamento de F\'{\i}sica, Universidade Federal de Santa Catarina, Trindade 88040-900, Florian\'{\o}polis, SC, Brazil\\
  $^9$ Instituto Milenio de Astrof\'{\i}sica, Santiago, Chile\\
  $^{10}$ Vatican Observatory, V00120 Vatican City State, Italy}

\date{Accepted XXX. Received YYY; in original form ZZZ}

\pubyear{2018}

\begin{document}
\label{firstpage}
\pagerange{\pageref{firstpage}--\pageref{lastpage}}
\maketitle

\begin{abstract}
Red clump stars are commonly used to map the reddening and morphology of the inner regions of the Milky Way. We use the new photometric catalogues of the VISTA Variables in the V\'ia L\'actea survey to achieve twice the spatial resolution of previous reddening maps for Galactic longitudes $\rm -10^{\circ}<l<10^{\circ}$ and latitudes $\rm -1.5^{\circ}<b<1.5^{\circ}$. We use these de-reddened catalogues to construct the $\rm K_{s}$ luminosity function around the red clump in the Galactic plane. We show that the secondary peak (fainter than the red clump) detected in these regions does not correspond to the bulge red-giant branch bump alone, as previously interpreted. Instead, this fainter clump corresponds largely to the over-density of red clump stars tracing the spiral arm structure behind the Galactic bar. This result suggests that studies aiming to characterise the bulge red-giant branch bump should avoid low galactic latitudes ($\rm |b|< 2^{\circ}$), where the background red clump population contributes significant contamination. It furthermore highlights the need to include this structural component in future modelling of the Galactic bar. 
\end{abstract}




\section{Introduction}
\label{sec:intro}

The Milky Way (MW) is the only galaxy in the Universe whose components can all be resolved into individual stars. As such it is an ideal, and indeed, unique laboratory to investigate the details of the processes behind formation and evolution of a disc galaxy \citep{Bland-Hawthorn+16}. As a result, the need to obtain a large-scale empirical description of the MW has defined the ambitious requirements of large photometric and spectroscopic surveys during the last decade. 

In recent years, and in particular with the availability of the near-IR photometry from the VISTA Variables in the Via Lactea  (VVV) ESO public survey, red clump (RC) stars have become one of the most used tool to investigate the extinction properties and the structure of the inner Galaxy. The RC can be identified easily in the observed colour-magnitude diagram (CMD) of stellar populations and their absolute magnitude is well defined by theoretical models \citep[see][for a recent review]{girardi+16}. For a given Galactic stellar population, such as that of the Milky Way bulge (MWB), the mean magnitude of its RC stars is affected by interstellar extinction, distance, and to a minor extent by changes in the stellar population properties such as age and metallicity. The mean metallicity of the MWB is well known to vary as a function of Galactic latitude, going from near-Solar metallicity at low latitudes to $\rm [Fe/H] \sim -0.40$ dex in the outermost regions. The change in mean metallicity as a function of Galactic longitude is much smaller, if at all present, within $\rm -10^{\circ}<l<10^{\circ}$. On the other hand, although the majority of MWB stars are known to be old ($\sim$10 Gyr), it remains to be understood if there is a significant fraction of young stars ($<$5 Gyr), specifically at the metal-rich end \citep[see][and references therein]{alvio+18}. However, the MWB age-metallicity relation recently found by \citet{bernard+18}, that favours a wide range of ages for metal-rich stars, can be used to show that in this case (for the well known spatial variations in mean metallicity) the mean age of the MWB could spatially vary from 10 Gyr in the outer regions to up to 6 Gyr closer to the plane. Even when these maximum population gradients are considered, the theoretical mean magnitude variation of the RC would be of $<0.10$ magnitudes \citep[see Fig.~6 in][]{salaris+02}, which is equivalent to a distance uncertainty of $\sim300$ pc at 8 kpc. It is for this reason that identification and mapping of the mean magnitude of the RC can be safely used both to correct for interstellar reddening as well as to obtain the distance distribution of MWB stars for a given line-of-sight (LOS). 

Since the study of \citet{stanek+94}, most structural studies of the MWB based on the RC follow a similar procedure that includes the construction of the luminosity function (LF) from the dereddened CMD and the fitting of the underlying red giant branch (RGB) stars using a polynomial or exponential function. The contribution of the RC in the fit is then parametrised using a Gaussian function. However, an additional, secondary peak in the LF (fainter than the RC) was first identified (but left out of the fit) by \citet[][see their Fig. 3]{nishi+05} at latitudes $\rm b=1^{\circ}$. The secondary peak was then included in the RC parametrisation adding a second Gaussian fit to OGLE-III data by \citet{nataf+11} at higher Galactic latitudes but lower than those where the double RC from the X-shaped bulge becomes important (i.e. at $\rm |b|<5^{\circ}$).  \citet{nataf+11} interpreted the secondary peak as the MWB red giant branch bump (RGBB) which was followed by a more detailed analysis of its properties by \citet{nataf+13}, who found that the RGBB was 0.10 mag brighter than expected therefore suggesting an anomalous helium enhancement to reconcile this difference. Later studies have included the two-Gaussian fit regardless of the region being studied and without further discussion on the nature of the secondary peak (relying on a RC+RGBB nature). \citet{gonzalez+11c} used the RC to trace the bar properties at latitudes $\rm |b|<2^{\circ}$ (the same region studied in \citet{nishi+05}), showing that the secondary peak does not follow the mean magnitude nor the density variations of the RC as a function of Galactic longitude. 

In this letter we use new VVV catalogues based on PSF photometry (Alonso-Garc\'{\i}a et al. 2018) to produce a high-resolution ($\rm 1\arcmin\times1\arcmin$) reddening map of the MWB regions between $\rm -10^{\circ}<b<+10^{\circ}$ and $\rm -2^{\circ}<b<+2^{\circ}$. We use the PSF catalogues and reddening map to show that the secondary peak in these in-plane regions is tracing the spiral arm structure of the disc behind the Galactic bar. 

\section{Observations}

Observations for the VVV survey were taken with the 4m VISTA telescope located in ESO Cerro Paranal Observatory in Chile. The VIRCAM camera installed on it contains 16 detectors, which provide near-infrared, wide, sparsely-sampled images of the sky known as \textit{pawprints}, covering a total area of $0.6$~deg$^2$ with a resolution of 0.34$\arcsec$ per pixel \citep[][]{minniti+10}. Combining a mosaic of six partially superimposed pawprint exposures results in a contiguous coverage of the so-called \textit{tile}, an image of $1.64$~deg$^2$ field of view \citep{saito+11}). Images in Z, Y, J, H, and $\rm K_s$ are processed and reduced using the VISTA Data Flow System \citep[VDFS][]{ emerson+04, irwin+04}, developed at the Cambridge Astronomical Survey Unit (CASU). VDFS also produces catalogues providing the positions of the objects detected, along with their aperture photometry. 

Since we are concentrating our efforts on the inner regions of the VVV survey  ($\rm -10^{\circ}<b<+10^{\circ}$ and $\rm -2^{\circ}<b<+2^{\circ}$), crowding and reddening effects are an important factor to take into account. Although the quality of the catalogues provided by CASU are of excellent quality and suitable for immediate science, the reliability of a study of the secondary peak strongly depends on the completeness and photometric accuracy at these faint magnitudes.  \citet{valenti+16} circumvent this issue by using VVV PSF catalogues for J, and $\rm K_s$ that outperform those based on aperture photometry in high-crowding areas. The completeness of those PSF catalogues in the bulge regions that are relevant for this work is $\sim80\%$ at a magnitude $K_s$=14, which is a factor two higher than those produced using aperture photometry.

\begin{figure*}
\centering
\includegraphics[width=15cm,angle=0,trim=0.0cm 0.8cm 0.0cm 1.8cm, clip=true]{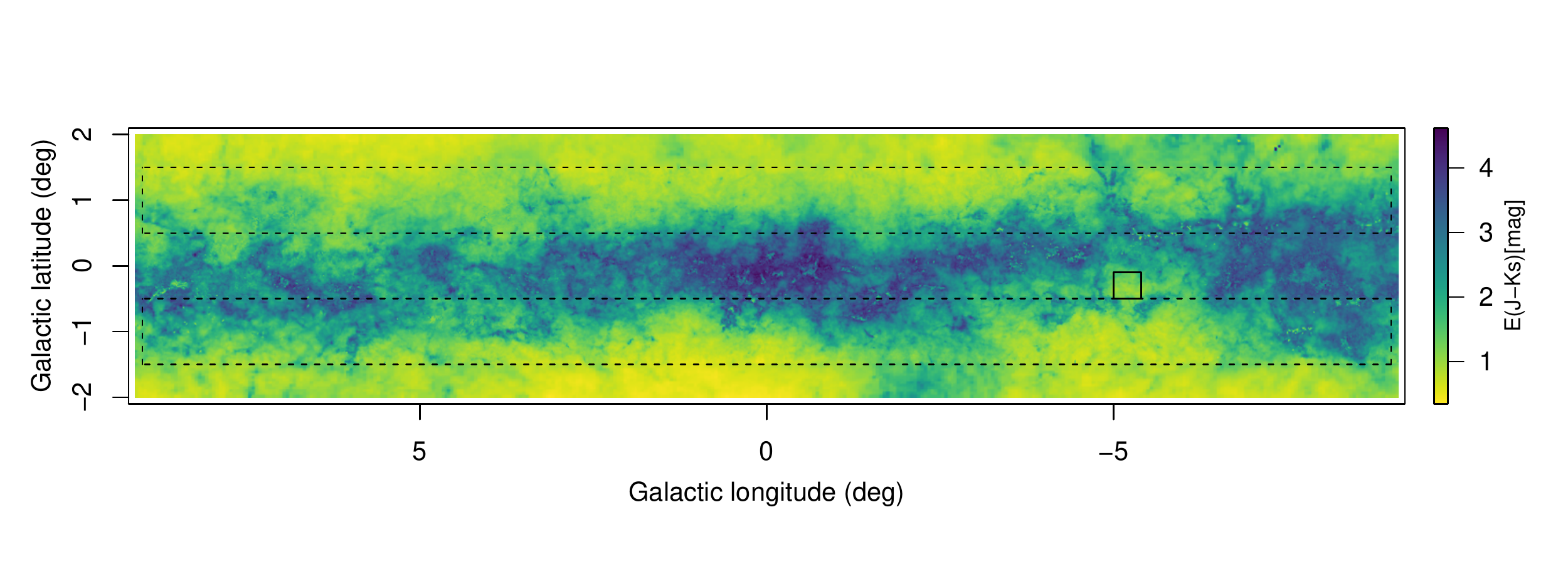}\\
\caption{Reddening map of the in-plane regions of the MWB, with a $\rm 1\arcmin$ spatial resolution, for Galactic longitudes $\rm -10^{\circ}<b<+10^{\circ}$ and latitudes $\rm -1.5^{\circ}<b<+1.5^{\circ}$. Dashed boxes mark the regions analysed in this work. A particularly low extinction window is also marked as a solid line box. The complete map, including the outer regions presented in \citet{gonzalez+12}, is available via an interpolation code from http://mill.astro.puc.cl/BEAM.}
\label{windows}
\end{figure*}

Further details on the PSF photometric catalogues is provided in a separate article (Alonso-Garc\'{\i}a et al. 2018). Here it is sufficient to mention that the PSF photometry is performed on the individual chips of the pawprint images in the $\rm J$ and $\rm K_s$ filters, using the DoPHOT software package \citep{Schechter+93, alonso-garcia+12}. We used the average magnitudes of two epochs per filter, weighted according to their reported uncertainties. Positions from the detected objects were astrometrised using the information provided by CASU, and the photometry calibrated into the VISTA photometric system applying a magnitude offset to every chip after direct comparison, in a given chip, of the brightest stars in common with the CASU catalogs. Individual chips were cross-correlated using the STILTS package (Taylor 2006) to individual tile fields first, and after that to all the region of interest. 

\section{The high-resolution reddening map of the inner MWB}

\subsection{The RC method to measure reddening}

In this work we look to improve the resolution of the reddening maps from \citet{gonzalez+12} ($\rm 2\arcmin \times 2\arcmin$) near to the Galactic plane ($\rm |b|<1.5^{\circ}$) by using deeper, more complete PSF photometry of the VVV data. Because the new PSF catalogues reach higher number counts in the RC region of the CMD, we are able to accurately detect the RC and measure its mean color in spatial bins of $\rm 1\arcmin \times 1\arcmin$. Therefore, we can calculate reddening values following the method described in \citet{gonzalez+11b} but improving the spatial resolution obtained in \citet{gonzalez+12} by a factor of two.

The selection of RC stars is made from the $\rm (J-K_s)$-$\rm K_s$ CMD. A Gaussian fit is applied to the RC $\rm (J-K_s)$ colour and the colour excess $\rm E(J-K_s)$ is calculated with respect to an intrinsic value of $\rm (J-K_s)=0.68$, as measured for de-reddened RC stars in Baade's Window \citep{gonzalez+12}. The RC star selection box is selected on tile-by-tile basis to ensure that the RC region is correctly mapped, accounting for the differences in reddening and distance in each LOS. Specifically, the $\rm K_s$ limits of the selection box are adjusted in each tile catalogue to account for the change in magnitude that follows the orientation of the Galactic bar. The blue colour limit is visually selected for each tile to minimize contamination from the disc main sequence. Finally, the red limit in the selection box is fixed to $\rm (J-K_s)=5.5$~mag as determined by the highest colour excess that can be traced by our catalogues. This limiting colour excess is determined by the limit magnitude of the J-band catalogues. The uncertainties in the $\rm E(J-K_s)$ values are dominated by the remaining differential reddening from variation within  spatial scales smaller than $\rm 1\arcmin$ and can reach values as high as $\rm \sigma E(J-K_s)\sim$ 0.20 magnitudes in the innermost regions. We quantify this error using the dispersion of the Gaussian fit to $\rm (J-K_s)$ for each bin as a proxy for the residual differential reddening. The residual reddening can be estimated by deconvolving the dispersion of the Gaussian fit in each bin with the intrinsic colour width of the RC in Baade's window, estimated to $\rm \sigma=0.10$ mag from de-reddened RC in Baade window \citep[as in][]{gonzalez+12}).

\begin{figure}
\centering
\includegraphics[width=7.6cm,angle=0,trim=0.0cm 0.2cm 0.0cm 1.8cm, clip=true]{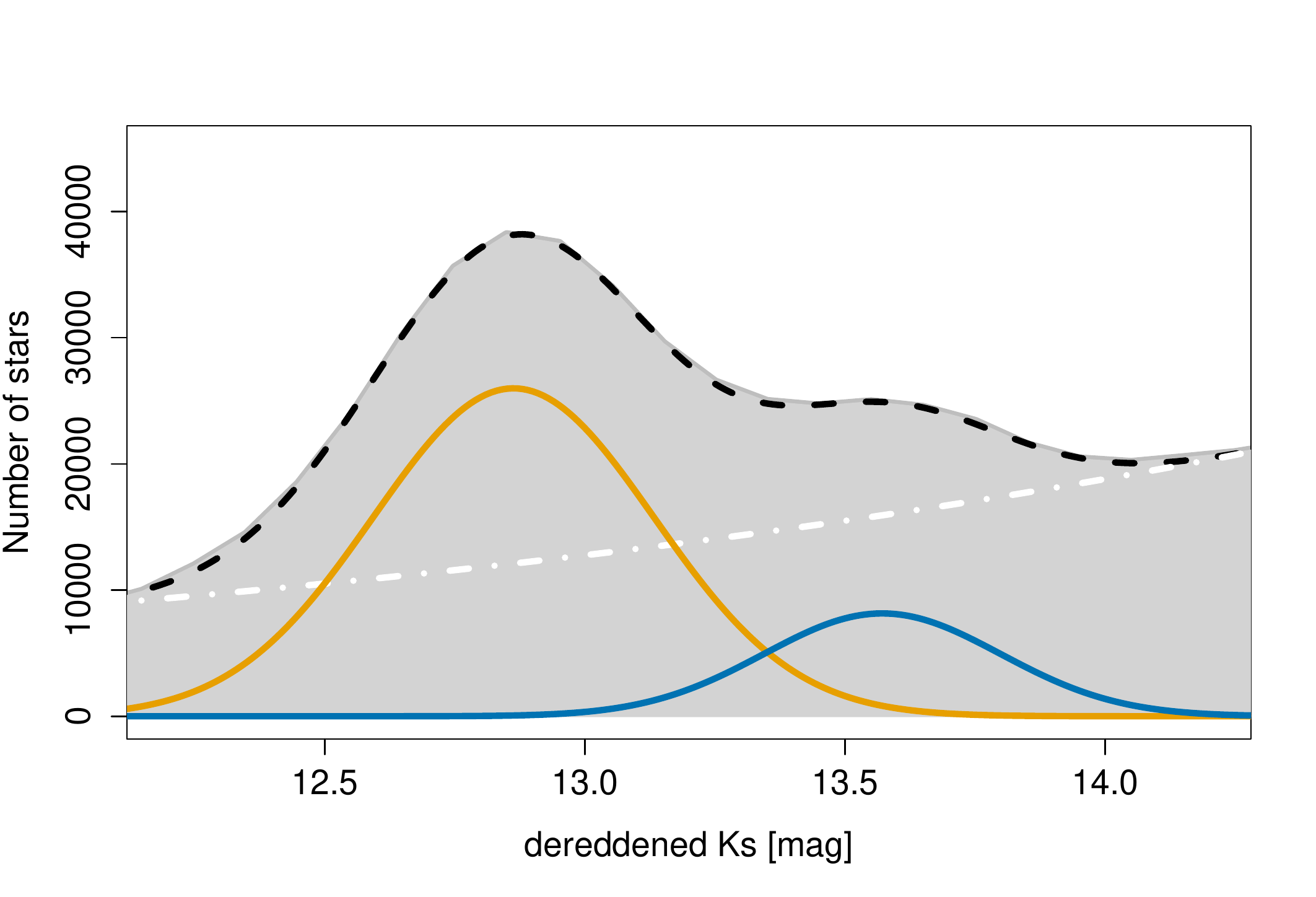}\\
\caption{$\rm K_{s,0}$ LF of a 1 sq. deg. field centred at $\rm (l,b)=(0^{\circ},-1^{\circ})$ (grey shaded area) and the best-fitting exponential+double Gaussian function (dashed black line). The exponential function for the RGB (white dash-dotted line) as well as a Gaussian function main RC (orange solid line) and the secondary peak (blue solid line) are also shown.}
\label{twoclumps}
\end{figure}

The improved resolution of the maps at low latitudes allows us to trace the fine structure of the dust features that can be seen in the $\rm K_s$ band images of a given region. The new reddening map allows to perform a better study of the CMD, providing reddening corrections for sources that are closer to the optimal value that would be measured from the individual source itself. This is particularly critical in the inner regions where the amount of reddening changes over very small spatial scales. However, one limitation of these 2D maps is that, by definition, their reddening values correspond to the integrated absorption along the line-of-sight at the mean distance of the observed MWB population. 

The applications of this reddening map range from the study of stellar populations such as age determinations via turn-off fitting, structure from the construction of the LF and fitting of the RC, as well as photometric first guess stellar parameters for spectroscopic abundance analysis. It can also be used to identify and explore new regions of low reddening in the VVV area \citep{minniti+18}. Here we identify a new low-reddening window in the Galactic plane located at $\rm (l,b)=(-5.2^{\circ},-0.3^{\circ})$ (see Fig.~\ref{windows}). This half-degree diameter window (VVV WIN 1733-3349) reaches $\rm E(J-K_s)$ values nearly 3 magnitudes lower than the surrounding field. We suggest this window to be used in future studies as a reference field for the inner MWB, similarly to how Baade's window has been used for the outer regions, where future multi-band surveys with different sensitivities can cross-calibrate their measurements.

\subsection{BEAM-II: Bulge Extinction And Metallicity calculator II}

In \citet{gonzalez+12} we presented a web application named Bulge Extinction And Metallicity calculator (BEAM) where the user can retrieve values for $\rm E(J-K_s)$ and photometric metallicities \citep[from][]{gonzalez+13} for a given field or a catalogue of coordinates. We have now prepared a new BEAM code, named BEAM-II, in R\footnote{https://cran.r-project.org} that includes the improved reddening values presented here. The code can run on any machine with a working R installation and is no longer limited to short catalogues as in the HTML-based BEAM applet. 

The BEAM-II code is based on \textit{interp} routine of the \textit{akima} package, that performs a linear interpolation of the four neighbouring points to the input coordinates in the reddening grid. The grid also includes uncertainty estimations  and the mean photometric metallicity ([Fe/H]) for the input coordinates based on interpolation of the maps from Gonzalez et al. (2013). We note that the metallicity maps are only available for the regions $\rm |b|>3^{\circ}$ as described in \citet{gonzalez+13}. 

\section{The structure behind the bar}
\begin{figure*}
\centering
\includegraphics[width=17cm,angle=0]{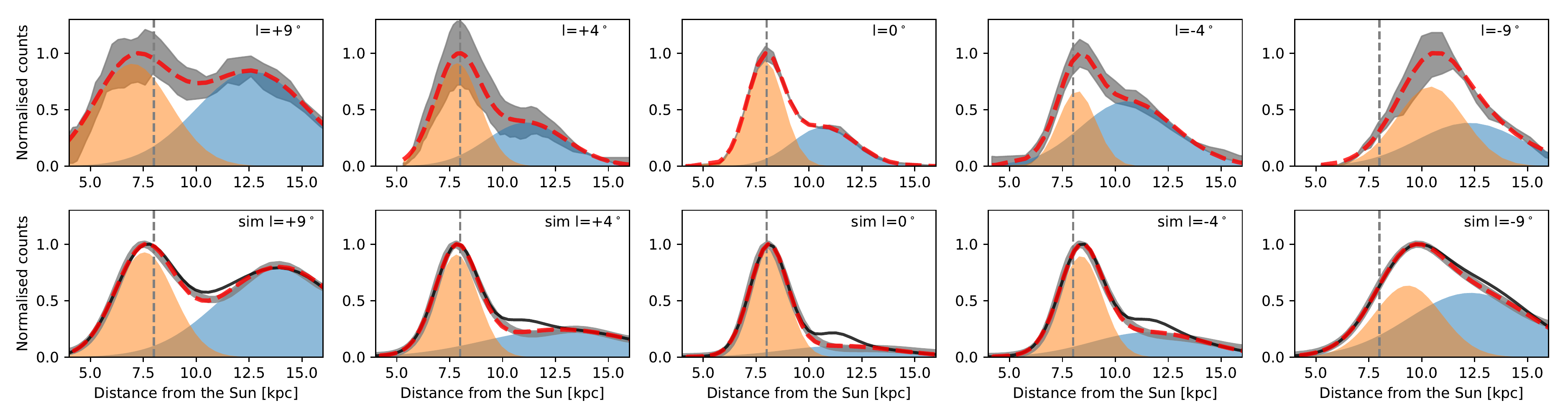}\\
\caption{Distance distribution of RC stars at Galactic latitude $\rm |b|=1^{\circ}$ (upper panel) and a range of longitudes (shown in each panel) from VVV (top row) and from the N-body simulation from \citet{gardner+14} (bottom row). Observed distances are obtained assuming all stars in the RGB-subtracted LF to be RC stars and the grey shaded area denotes the difference between positive and negative latitudes. The best-fit Gaussian functions to the mean distance distribution for the main RC (orange) and secondary peak (blue) are also show. The red dashed lines show the resulting best fit distribution to the observations. In the simulations the black solid line shows the resulting distribution after including a RGBB-like Gaussian to the LF. The vertical grey dashed lines marks a distance of 8 kpc in all panels for reference.}
\label{lf_clump}
\end{figure*}

\begin{figure}
\centering
\includegraphics[width=8.0cm,angle=0]{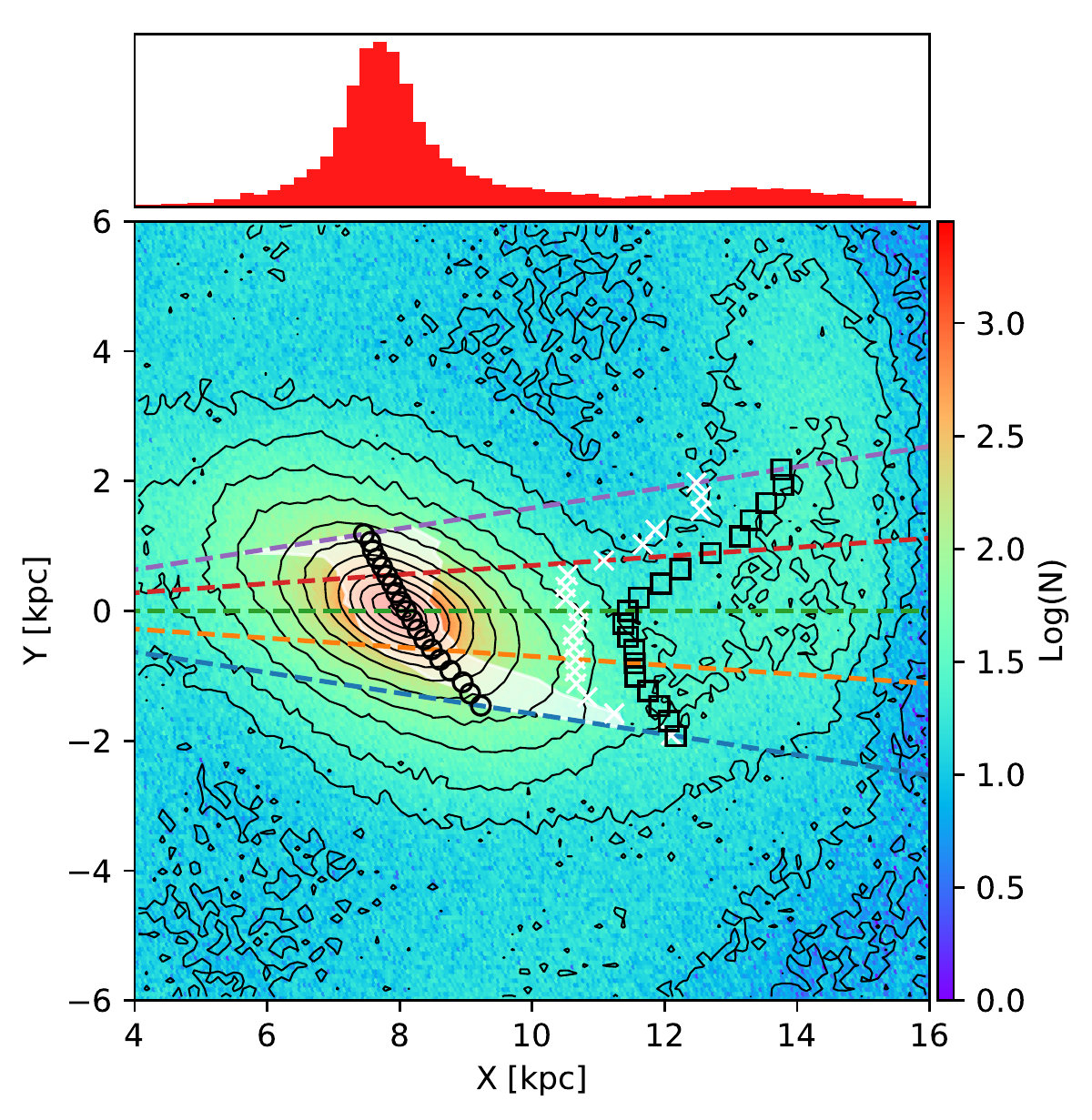}\\
\caption{Projected density of stars in simulation R1 from \citet{debattista+05}, scaled to the MW as in \citet{gardner+14}, restricted to $z=\pm0.4$ kpc. The mean distance to the near (circles) and far (squares) structures along the LOS distance distribution of the simulation in $\rm 1^{\circ}\times1^{\circ}$ bins are shown. The white shaded area corresponds to the $\rm 1\sigma$ around the mean of the Gaussian fit for RC1 in the MW observations. The mean distance to the secondary peak in the MW is shown with white crosses. Note that the fit in the observed case is also done in distance space, instead of magnitude as usually done in previous studies, under the assumption that all stars belong to the RC. LOS for l=-9$^{\circ}$, -4$^{\circ}$, 0$^{\circ}$, +4$^{\circ}$, and +9$^{\circ}$ used to construct the distance distributions of Fig.\ref{lf_clump} are also marked in the figure. The distance distribution for the line of sight l=+4$^{\circ}$ is shown in the upper panel to show the background over-density produced by the spiral arm structure.}
\label{projected}
\end{figure}

Thanks to the superior photometry of the VVV PSF catalogues used in this study, we can improve our previous measurements of reddening and then apply it to investigate the structure of the MWB and bar near to the Galactic plane. In particular, we focus the analysis on the secondary peak observed in the LF following-up on the results from \citet{gonzalez+11c}.  \citet{gonzalez+11c} suggested that the behaviour of the RGBB at $\rm b \sim 1^{\circ}$ is not consistent with a population feature like the RGBB as its mean magnitude had a different variation as a function of Galactic latitude than the RC. Note that a secondary, fainter peak is still detected at intermediate Galactic latitudes ($\rm b\sim4^{\circ}$) \citep[see Fig.~9 in][]{gonzalez+11b} but its mean magnitude and strength changes are correlated with those of the RC. This behaviour at intermediate latitudes is consistent with the expectations from the RC+RGBB (in terms of their mean magnitude changes), in agreement with the analysis of \citet{nataf+11}. Here we investigate the properties of the secondary peak in the Galactic plane $\rm |b|=\pm1^{\circ}$ using the improved photometric VVV catalogues and corresponding reddening maps.

We first construct the dereddened $\rm K_{s,0}$ LF in each tile applying a colour cut of $\rm (J-K_s)_0>0.30$ mag to limit contamination from the foreground disc. Figure~\ref{twoclumps} shows the best fit to the LF at $\rm (l,b)=(0^{\circ},-1^{\circ})$ using an exponential + double Gaussian function. The need for a double Gaussian in the fit is evident from inspecting Fig.~\ref{twoclumps}.  Starting from the hypothesis that, at these low latitudes, the secondary peak is not the RGBB but the RC of a background disc structure, we assume the entire RGB-free LF to be composed of RC stars and therefore following the LOS distance distribution of stars.

To evaluate this hypothesis, we subtract the best fit exponential function from the observed LF in each field, leaving only the main RC and secondary peak features (hereafter referred to as RC1 and RC2, respectively). We construct the LOS distance distribution bins of $\rm 1^{\circ}\times1^{\circ}$ by adopting an intrinsic magnitude of the RC of $\rm M_{K_s}(RC)=-1.61$ mag \citep{chen+17}. The resulting distance distributions for longitudes $\rm l=-9^{\circ}, -4^{\circ}, 0^{\circ}, +4^{\circ}$, and $\rm +9^{\circ}$ are shown in Fig.~\ref{lf_clump}. While the location of RC1 changes as a function of longitude as expected from the position angle of the bar, the location of RC2 moves in the opposite direction, increasing their separation towards positive longitudes. Similarly, their relative density depends on longitude, with a RC1/RC2 ratio that clearly increases towards $\rm l=0^{\circ}$.  Note that the density of RC1 and RC2 are similar at positive longitudes (see leftmost panel in Fig.~\ref{lf_clump}). Neither of these properties are consistent with the RGBB but, instead, they justify our assumption of the secondary peak being dominated by red clump stars from a structure located behind the bar, such as a spiral arm. 

To verify this conclusion we investigate if a spiral arm behind the bar would produce such a feature in the distance distribution by looking at an N-body simulation of a Milky Way-like galaxy from \citet{debattista+05}. Simulation R1 from \citet{debattista+05} is a disc galaxy that forms a B/P bulge via buckling instabilities of the bar. Because of its similarity with the Milky Way, R1 is part of the Gaia challenge \footnote{http://astrowiki.ph.surrey.ac.uk/dokuwiki/doku.php} and, among other studies, it has been used to investigate the X-shape properties of the MW \citep{gardner+14, vasquez+13}. In this work we use R1, scaled to the Milky Way size as described in \citet{gardner+14}, to obtain the distance distribution of stars along the same LOS where we are observing the faint bump. To construct the distance distribution in a way that is comparable to those obtained using the RC magnitudes, we first translate the distances in the simulation to pseudo-RC magnitudes and convolve the magnitude distribution using a Gaussian of $\rm \sigma=0.20$ mag to account for the intrinsic width of the RC and residual differential reddening \citep[see for example][]{wegg-gerhard+13, gonzalez+11}. Finally, the pseudo-RC magnitude distribution of the simulation, now convolved with the observed RC distance uncertainty in magnitude space, are converted back to distance. 

The final LOS distances for the simulation are shown in the bottom row of Fig.~\ref{lf_clump}. The observed and simulated distance distributions behave very similarly in all the fields. In particular, we observe that the changes of the observed RC2 as a function of longitude resemble those of an over-density in the simulation, located behind the bar. The most important differences arise in the central fields at $\rm l=0^{\circ}$. In this particular LOS, the secondary peak could indeed include the RGBB as the density of MWB stars would we much higher than those of the background RC stars, as evidenced in the simulation (that is not affected by population effects) at $\rm l=0^{\circ}$ in Fig.~\ref{lf_clump}. We show this effect in the bottom panel of Fig.~\ref{lf_clump} when including a RGBB-like feature to the simulation by adding a simple Gaussian to the best fit distance distribution (in magnitude space before converting it back to distance). To produce the RGBB-like feature we assumed a simple Gaussian that is 0.71 magnitudes fainter than the RC, with 12\% the number density of RC stars and the same dispersion \citep{nataf+11}. This is not an attempt to model the RGBB itself but to show that indeed the inclusion of an RGBB-like feature produces a better match between the simulation and the observations at $\rm l=0^{\circ}$, while its contribution considerably decreases with respect to the background over-density as $\rm |l|$ increases. This further supports the interpretation of RC2 being dominated by the contribution of a structural feature different from the Galactic bar.

In order to identify the nature of the background structure in the simulation, we construct the projected density for particles restricted to the Galactic plane ($z=0.4$ kpc, equivalent to $\rm b=1.5^{\circ}$ at a distance of 16 kpc) which include the relevant regions for comparison to our VVV data. The projected density of the simulation is shown in Fig.~\ref{projected} together with the location of the observed and simulated mean distances shown in Fig.~\ref{lf_clump}. These mean distances are obtained from fitting a double Gaussian function to the LOS distance distributions. From Fig.~\ref{projected} it is clear that the origin of the background density peak is produced by the spiral arm structure. Because of the analogous behaviour of RC1 and RC2 to the bar+spiral arm in the simulation we conclude that RC2 traces the background spiral arm of the MW. 

It is important to note that the mean distance measured for background spiral does not match the actual location of the spiral due to projection effects, the merging of the bar and the spiral at its far end, and the fact that the RGBB should still be present in the observations. However, from the projected density of the simulation and the results from the Gaussian fit shown in Fig.~\ref{projected} we see that the simulated mean density matches fairly well the actual position of the spiral arm at positive latitudes. In the observations at this LOS, RC2 is located at $\sim12.7$ kpc from the Sun, which agrees with the expected location of Perseus arm in the Milky Way \citep[see Fig.~2 in][]{sanna+17}. However, this value can be affected by the adoption of an intrinsic magnitude of the RC different to that of the Galactic bar and by the potential presence of additional dust extinction behind the bar, as our 2D reddening maps are dominated by the dust lane in front of the bar \citet{minniti+14}. These effects must be taken into account when detailed maps and modelling of the spiral structure of the MW are done based on RC stars in future studies.

\section{Summary}

In this letter we used RC stars from the improved PSF photometric catalogues of the VVV survey in a stripe at $\rm b=-1^{\circ}$ to investigate the nature of the secondary peak, fainter than the RC in the LF, previously identified as the RGBB. We show that in these in-plane regions the secondary peak does not follow the bar position angle and density, as would be expected if it were associated with the same structure but different evolutionary phase (i.e. RGBB), confirming the results from \citet{gonzalez+11c}. Here we demonstrate, using a N-body simulation of a Milky Way-like galaxy, that the RC and the secondary fainter peak are consistent with the bar and the spiral arm behind the Galactic bar. 

From these results, we highlight the following:

\begin{itemize}
\item Studies attempting to characterise the RGBB of the MWB population should concentrate on Galactic latitudes between $\rm |b|=3^{\circ}$ and $5^{\circ}$ to avoid the effects of the background spiral arm at lower latitudes and of the two arms of the X-shaped bulge at latitudes larger than $\rm |b|\sim 6^{\circ}$. Note that \citet{nataf+11} analysed OGLE-III data that concentrates mostly at $\rm b>-2^{\circ}$ (see their Fig. 2). Therefore, we expect their conclusions to remain valid, but a revision of the influence of the spiral structure behind the bar on their innermost fields in their study is encouraged.
\item There is important structural information in the secondary peak that should not be neglected or confused as a pure population effect such as the RGBB. Models used to characterise the structure of the inner MW should take this feature into account, particularly in the LOS towards the far-end of the bar where the two structures merge.
\item We find that the structure is located approximately between 12.0-13.0 kpc from the Sun towards longitudes $\rm l=0^{\circ}$, possibly matching the location of the Perseus arm. Further modelling is required in order to map this structure in detail.

The fact that RC stars are tracing the spiral arm behind the galaxy presents an important opportunity for investigating the chemo-dynamical properties of the far side of the Galaxy with high number statistics. The Multi-Object Optical and Near-IR Spectrograph (MOONS) to be installed at the Very Large Telescope in late 2020 will provide radial velocities and chemistry for thousands of stars in this region. This data coupled with the positional information from VVV data will allow us to map the backyard of the Milky Way in unprecedented detail.
\end{itemize}

\section*{Acknowledgements}

OAG gratefully acknowledges support from the ESO Scientific Visitor Programme in Chile under which this work was completed. VPD is supported by STFC Consolidated grant \#~ST/R000786/1. J.A-G. acknowledges support by FONDECYT Iniciaci\'on 11150916 and by the Chilean Ministry of Economy through ICM grant IC120009 awarded to the Millennium Institute of Astrophysics (MAS). RKS acknowledges support from CNPq/Brazil through projects 308968/2016-6 and 421687/2016-9. MH acknowledges support the BASAL Center for Astrophysics and Associated Technologies (CATA) through grant PFB-06.




\bibliographystyle{mnras}
\bibliography{mybiblio_rev_full}








\bsp	
\label{lastpage}
\end{document}